\documentclass[preprint,12pt]{elsarticle}
\usepackage{amssymb}

\journal{Optics Communications}

\input{tcilatex}

\begin{document}

\begin{frontmatter}



\title{Optical properties in a two-dimensional quantum ring: Confinement potential and Aharonov-Bohm effect}


\author{Shijun Liang$^{\ast }$\footnote{E-mail: shijun\_liang@163.com }, Wenfang Xie$^{\ast }$}

\address{\small ($^{\ast }$Department of
Physics, College of Physics and Electronic Engineering, Guangzhou
University,
Guangzhou 510006, P.R. China)}

\begin{abstract}
Optical properties of a two-dimensional quantum ring with pseudopotential in the presence of an external magnetic field and magnetic flux have been theoretically investigated. Our results show that both of the pseudopotential and magnetic field can affect the third nonlinear susceptibility and oscillator strength. In addition, we found that the oscillator strength and the absolute value of the resonant peak of the linear, non-linear and total absorption coefficient vary periodically with magnetic flux, while the resonant peak value of the linear, non-linear and total refractive index changes decreases as magnetic flux increases.
\end{abstract}


\end{frontmatter}



\section{\textbf{Introduction}}
Since 1970s Scientific research into electronic structure had been devoted to two-dimensional structure --- quantum wells [1,2], the new and unusual properties of quasi-two-dimensional systems, which promise applications mostly in electronics and opto-electronics, have attracted the attention of many researchers [3-10]. This in turn has resulted in a rapid development of production technology and extensive research. This rapid progress in technology made it possible to create the quantum wire and quantum dot. Due to much more particular properties of electron confined in a quantum dot, a lot of studies on quantum dot have been done experimentally and theoretically [11-20].
   One of the subjects concerned with quantum dots is to investigate their optical properties, especially nonlinear optical properties. Over the last decades, researchers have reported the linear and nonlinear optical properties of semiconductor quantum dots [21-30]. For instance, G. Rezaei, M. R. K. Vahdani and B. Vaseghi [31] studied nonlinear optical properties of a hydrogenic impurity in an ellipsoidal finite potential quantum dot, their results show that the light intensity, size and geometry of the dot and aluminium concentration have a great influence on the absorption coefficient and refractive index changes of the dot. H. A. Sarkisyan $et$ $ al$ [32] presented indirect transitions in thin films due to the Coulomb interactions between electrons and  the frequency dependence as well as dependence on the concentration of conductivity electrons and thickness of the film has been obtained. Karabulut [33] reported laser field effect on the nonlinear optical properties of a square quantum well under the applied electric field, the results show that the laser field considerably affects the confining potential of the quantum well and the nonlinear optical properties.

     As we know, modern electronic and optoelectronic devices can be nanometric dimensions where microscopic details can not be treated in an effective way, atomistic approaches become necessary for modelling structural, electronic and optical properties of such nanostructured devices, and the pseudopotential plays a important role in the studies of semiconductor low-dimensional structures. Researchers often use the Pseudopotential approaches to theoretically obtain  some important information about electronic structures in semiconductor [34-43]. In addition, the pseudopotential was applied to interpreting some results from experiments with great success [44-46].
  And it is well-known that Aharonov-Bohm effect actually refers to the quantum mechanical phase of the
wave function which is not a physical observable. We can acquire the AB phase by a charged carrier which traverses a region where magnetic field doesn't exist. Since Aharonov and Bohm [47] provided the interpretation for Aharonov-Bohm effect in 1959, a lot of studies [48-57] have been reported about this topic. Those papers covered many properties of low-dimensional semiconductor structures, some of which is focused on the optical properties of low-dimensional semiconductors [58-61]. And most researchers focused their studies on the quantum ring and results from calculations or experiments indicate that optical properties of quantum ring are strongly affected by the Aharonov-Bohm effect.
   Recently, some new nanostructures, such as antidots, have attracted much attention. And researchers have reported in a lot of literatures [62-65] about these structures. It would be very interesting if we can investigate a quantum ring with pseudopotential. For this purpose, we will focus on studying effects of an external magnetic field, magnetic flux quantum, pseudopotential on the optical properties of a two-dimensional quantum ring in the present paper. The paper is
organized as follows: in section 2 we describe the model and theoretical
framework, section 3 is dedicated to the results and discussions, and
finally, our conclusions are given in section 4.

\section{\textbf{Model and calculations}}

\subsection{Electronic state in a two-dimensional quantum ring with the pseudoharmonic potential}

Consider a two-dimensional quantum ring with pseudopotential, total Hamiltonian of the system with a uniform magnetic field $\mathbf{B}$ and AB field applied simultaneously in the z-direction can be written as

\begin{equation}
H=\frac{1}{2m_{e}^{\ast }}\left[ \mathbf{p+}\frac{e}{c}\mathbf{A}\right]
^{2}+V(\mathbf{r}),
\end{equation}
In Eq.(1), $m_{e}^{\ast }$ is electronic effective mass, $e$ is the electron
charge, $c$ is the speed of light, $\mathbf{A}$ is a sum of two terms, $\mathbf{A}=\mathbf{A_{1}}+\mathbf{A_{2}}$ such that $\nabla\times\mathbf{A_{1}}=\mathbf{B}$ and $\nabla\times\mathbf{A_{2}}=0$ for $r\neq0$, where $\mathbf{B}$ denotes magnetic field. $V(\mathbf{r})$ is the pseudopotential given as follows [65]
\begin{equation}
V(\mathbf{r})=V_{0}(\frac{r}{r_{0}}-\frac{r_{0}}{r})^{2},
\end{equation}

where $V_{0}$ denotes the confinement strength on the two-dimensional electron gas
and $r_{0}$ is the zero point of the potential. With the
gauge ${\mathbf{A}}{_{1}^{\varphi}}=\frac{Br}{2}$ and ${\mathbf{A}}{_{2}^{\varphi}}=\frac{\Phi_{AB}}{2\pi r}$, this is, $\mathbf{A}=(0,\frac{Br}{2}+\frac{\Phi_{AB}}{2\pi r},0)$, the Schr$\ddot{o}$dinger equation in cylindrical coordinates can be written as
\begin{equation}
-\frac{\hbar^{2}}{2m^{\ast}_{e}}[\frac{1}{r}\frac{\partial}{\partial r}(r\frac{\partial}{\partial r})+\frac{1}{r^{2}}(\frac{\partial}{\partial \varphi}
+i\frac{\Phi}{\Phi_{0}})^{2}]\Psi\nonumber \\
+i\frac{\hbar\omega_{c}}{2}(\frac{\partial}{\partial\varphi}+i\frac{\Phi_{AB}}{\Phi_{0}})\Psi+\frac{m^{\ast}_{e}\omega_{c}^{2}r^{2}}{8}\Psi+V(\mathbf{r})\Psi = E\Psi,
\end{equation}
where $\omega_{c} = eB/m^{\ast}_{e}c$ is the cyclotron frequency, $\Phi_{0}=\frac{hc}{e}$ is magnetic flux quantum.
The wave functions and the energy spectrum of an electron confined in a quantum ring can be obtained by solving the Schr$\ddot{o}$dinger equation.
\begin{equation}
\Psi(r,\varphi) = Nr^{|\xi|}e^{-\delta r^{2}}F(-n,|\xi| + 1;\delta r^{2})e^{im\varphi},
\end{equation}

\begin{equation}
E_{nm}(\beta) = \hbar(n+\frac{|\xi|+1}{2})\sqrt{\omega_{c}^{2}+\frac{8V_{0}}{m^{\ast}_{e}r_{0}^{2}}}+\frac{(m+\alpha)\hbar\omega_{c}}{2}-2V_{0}.
\end{equation}
With \begin{equation}
\xi = \sqrt{(m+\alpha)^{2}+\frac{2m^{\ast}_{e}V_{0}r^{2}_{0}}{\hbar^{2}}},
\end{equation}
and
\begin{equation}
\delta = \sqrt{\frac{e^{2}B^{2}}{4\hbar^{2}c^{2}}+\frac{2m^{\ast}_{e}V_{0}}{\hbar^{2}r^{2}_{0}}}.
\end{equation}
Where $N$ is the normalization constant. $F(a,b;x)$ is the confluent hypergeometric function. $n=0,1,2\cdots,$ is main quantum number, $m=0,\pm1,\pm2,\cdots,$ is magnetic quantum number. $\alpha=\frac{\Phi_{AB}}{\Phi_{0}}$ is dimensionless measure of magnetic flux $\Phi_{AB}$ which is created by a solenoid inserted into center of the quantum ring.

\subsection{\textbf{Calculation of the linear and the third-order nonlinear
optical absorption coefficient and refractive index changes}}

We employ compact-density approach to calculate the absorption coefficient and the changes of the
refractive index for a two-dimensional quantum ring structure. Suppose the system is excited by an electromagnetic field as
\begin{equation}
E(t) = E_{0}\cos(\omega t) = \widetilde{E}e^{i\omega t} + \widetilde{E}%
e^{-i\omega t}.
\end{equation}

The electronic polarization $\mathbf{P}(t)$ and susceptibility $\chi(t)$ are
defined by the dipole operator $M$, and the density matrix $\rho$,
respectively
\begin{equation}
\mathbf{P}(t) = \epsilon_{0}\chi(\omega) \widetilde{E}e^{i\omega t} +
\epsilon_{0}\chi(-\omega)\widetilde{E}e^{-i\omega t} = \frac{1}{V}Tr(\rho M).
\end{equation}
Where $\epsilon_{0}$ is the permittivity of free space, V denotes the volume
of the system. $Tr$ denotes the trace or summation over the diagonal
elements of the matrix $\rho M$. We can obtain the analytic expressions
[67] of the linear and the third-order nonlinear susceptibilities.

For the linear term
\begin{equation}
\epsilon _{0}\chi ^{(1)}(\omega )=\frac{\rho|M_{fi}|^{2}}{E_{fi}-\hbar \omega
-i\hbar \Gamma _{if}},
\end{equation}%

For the nonlinear term
\begin{equation}
\epsilon _{0}\chi ^{(3)}(\omega ) =-\frac{\rho|M_{fi}|^{2}\widetilde{E}^{2}}{
E_{fi}-\hbar \omega -i\hbar \Gamma _{if}}[\frac{4|M_{fi}|^{2}}{(E_{fi}-\hbar
\omega )^{2}+(\hbar \Gamma _{if})^{2}}  \nonumber \\
-\frac{(M_{ff}-M_{ii})^{2}}{(E_{fi}-i\hbar \Gamma _{if})(E_{fi}-\hbar
\omega -i\hbar \Gamma _{if})}],
\end{equation}%
where $\rho$ denotes the carrier density. $E_{fi}=E_{f}-E_{i}$ is the energy
interval of the two level system. $M_{fi}=e<\Psi _{f}|x|\Psi _{i}>$ is the
electric dipole moment of the transition from the $\Psi _{i}$ state to $\Psi
_{f}$ state. $\Gamma $ is the phenomenological operator. Non-diagonal matrix
element $\Gamma _{if}(i\neq f)$ of operator $\Gamma $, which is called as relaxation rate of $fth$ state, is the inverse of the relaxation time $T_{if}$, In our calculations $\Gamma
_{if}=1/T_{if}=1/0.2ps$ [68]. The susceptibility $\chi (\omega )$ is related to
the changes in the refractive index $\triangle n(\omega )/n_{r}$ and the
absorption coefficient $\alpha (\omega )$ as follows
\begin{equation}
\frac{\triangle n(\omega )}{n_{r}}=Re(\frac{\chi (\omega )}{2n_{r}}),
\end{equation}%
\begin{equation}
\alpha (\omega )=\omega \sqrt{\frac{\mu }{\epsilon _{R}}}Im(\epsilon
_{0}\chi (\omega )),
\end{equation}%
where $\mu $ is the permeability of the material, $\epsilon
_{R}=n_{r}^{2}\epsilon _{0}$ ($n_{r}$ is the refractive index) is the real
part of the permittivity.

The linear and third-order nonlinear absorption coefficients are obtained as
follows
\begin{equation}
\alpha ^{(1)}(\omega )=\omega \sqrt{\frac{\mu }{\epsilon _{R}}}\frac{%
\rho_{s}|M_{fi}|^{2}\hbar \Gamma _{if}}{(E_{fi}-\hbar \omega )^{2}+(\hbar \Gamma
_{if})^{2}}
\end{equation}%
\begin{eqnarray}
\alpha ^{(3)}(\omega ,I) &=&-\omega \sqrt{\frac{\mu }{\epsilon _{R}}}(\frac{I%
}{2\epsilon _{0}n_{r}c})\times \frac{\rho_{s}|M_{fi}|^{2}\hbar \Gamma _{if}}{%
(E_{fi}-\hbar \omega )^{2}+(\hbar \Gamma _{if})^{2}}[4|M_{fi}|^{2}-  \nonumber
\\
&&\frac{|M_{ff}-M_{ii}|^{2}[3E_{fi}^{2}-4E_{fi}\hbar \omega +\hbar
^{2}(\omega ^{2}-\Gamma _{if}^{2})]}{E_{fi}^{2}+(\hbar \Gamma _{if})^{2}}].
\end{eqnarray}%
Here $I$ is is the intensity of incident radiation. So, the total absorption coefficient $\alpha (\omega ,I)$ is given by
\begin{equation}
\alpha (\omega ,I)=\alpha ^{(1)}(\omega )+\alpha ^{(3)}(\omega ,I).
\end{equation}

The linear and the third-order nonlinear refractive index changes are
obtained as follows
\begin{equation}
\frac{\triangle n^{(1)}(\omega)}{n_{r}} = \frac{\rho_{s}|M_{fi}|^{2}}{%
2n_{r}^{2}\epsilon_{0}}\frac{E_{fi}-\hbar\omega}{(E_{fi}-\hbar\omega)^{2}+(%
\hbar\Gamma_{if})^{2}},
\end{equation}

and
\begin{eqnarray}
\frac{\triangle n^{(3)}(\omega )}{n_{r}} &=&-\frac{\rho_{s}|M_{fi}|^{2}}{%
4n_{r}^{3}\epsilon _{0}}\frac{\mu cI}{[(E_{fi}-\hbar \omega )^{2}+(\hbar
\Gamma _{if})^{2}]^{2}}  \nonumber \\
&&\times \lbrack 4(E_{fi}-\hbar \omega )|M_{fi}|^{2}-\frac{%
|M_{ff}-M_{ii}|^{2}}{(E_{fi})^{2}+(\hbar \omega )^{2}}((E_{fi}-\hbar \omega )
\nonumber \\
&&\lbrack E_{fi}(E_{fi}-\hbar \omega )-(\hbar \Gamma _{if})^{2}]-(\hbar
\Gamma _{if})^{2}(2E_{fi}-\hbar \Gamma ))].
\end{eqnarray}%
Therefore, the total refractive index change $\triangle n(\omega)/
n_{r}$ can be written as
\begin{equation}
\frac{\triangle n(\omega )}{n_{r}}=\frac{\triangle n^{(1)}(\omega )}{n_{r}}+%
\frac{\triangle n^{(3)}(\omega )}{n_{r}}.
\end{equation}

\subsection{Oscillator strength}

The oscillator strength of a transition is a dimensionless number which is useful for comparing different transitions. And it is a very important physical quantity in the study of the optical properties which are related to the electronic dipole-allowed transitions. Generally, the oscillator strength $P_{fi}$ is defined as
\begin{equation}
P_{fi}=\frac{2m_{e}^{\ast}}{\hbar^{2}}E_{fi}|M_{fi}|^{2}.
\end{equation}
\section{\textbf{Results and Discussions}}
\subsection{Effects of magnetic field and pseudopotential on the third-order nonlinear susceptibility}
Our calculations are performed for $GaAa/Al_{x}Ga_{1-x}As$ quantum dot. The parameters chosen in this work are the followings:
$m_{e}^{\ast} = (0.067 + 0.083x)m_{0}$, where $m_{0}$ is the free electron mass and $x=0.3$. $\rho_{s} = 5\times 10^{16}cm^{-3}$. In this section, we set magnetic flux to zero. The third-order susceptibility as a function of the photon energy with $r_{0} = 4 nm$ and $V_{0} = 350 meV$ for three different magnetic field values, is shown in Fig. 1. From this figure, we can find that the resonance peak decreases as the magnetic field increases, and that the peak position moves towards higher energies. Also, we can observe that the external magnetic field has a weak effect on the third-order susceptibility in two-dimensional quantum ring. In order to give an explanation for these behaviors, energy differences and the product of geometric factor as a function of the magnetic field with $r_{0} = 4 nm$ and $V_{0} = 350 meV$, are plotted in Fig. 2 and Fig. 3. It can be seen that the energy intervals slightly increase and the product of geometric factor, $\alpha_{03}\alpha_{32}\alpha_{21}\alpha_{10}$, considerably decreases when increasing the external magnetic field. Therefore, both two trends can contribute to reducing the third-order susceptibility. Moreover, in physical statement, By the increasing of the field the localization increases, resulting to the increase of the overlap integral. That is why the peak value of the third-order susceptibility decreases. In Fig. 4, we plot the third-order susceptibility as a function of the photon energy with $B = 1 T$ and $V_{0} = 350 meV$ for three different $r_{0}$ values. We can clearly observe that the resonance peak of third-order susceptibility increases dramatically as $r_{0}$ increases. Also, it is found that a red shift occurs in the resonance peak. The physical origins are shown in Fig. 5 and Fig. 6. As can be seen from Fig. 5 and Fig. 6, the energy differences considerably decrease for smaller $r_{0}$ and the product geometric factor $\alpha_{03}\alpha_{32}\alpha_{21}\alpha_{10}$, sharply increases as $r_{0}$ increases, both of which enhance the resonance peak value. So, we can draw a conclusion that the third-order susceptibility is strongly dependent on the $r_{0}$. The third-order susceptibility as a function of the photon energy with $B = 1 T$ and $r_{0} = 4 nm$ for three different $V_{0}$ values, has been presented in Fig. 7, to study the effect of $V_{0}$ on the third-order susceptibility. This figure clearly exhibits that increasing $V_{0}$ leads to the increment in resonance peak of third-order susceptibility. Meanwhile, we can also see that the peak position shifts towards higher energies with increased $V_{0}$. Next, we illustrate these behaviors by Fig. 8 and Fig. 9. We can find that from Fig. 8 energy intervals increase with $V_{0}$. And in Fig. 9, product of the geometric factor is also enhanced with $V_{0}$. However, the magnitude of the increment in the product of the geometric factor is much bigger than that in energy differences. Hence, the third-order susceptibility is enhanced. In Fig. 10, we demonstrate the third-order susceptibility of two-dimensional $Al_{x}Ga_{1-x}As$ pseudodot system as a function of the photon energy with $B = 1 T$, $r_{0} = 4 nm$ and $V_{0} = 350 meV$ for three different aluminium concentration $x$ values. From this figure, we can observe that aluminium concentration plays an important role in the third-order susceptibility of two-dimensional $Al_{x}Ga_{1-x}As$ pseudodot system, the peak value increases as aluminium concentration $x$ increases. In addition, it can be seen that the resonance peak moves to lower energies. Finally, in Fig. 11, we show the third-order susceptibility as a function of the photon energy with $B = 1 T$, $r_{0} = 4 nm$ and $V_{0} = 350 meV$ for three different $\tau$ (relaxation time) values. The effect of relaxation time on the third-order susceptibility is obvious. The longer the relaxation time is, the bigger the peak value of third-order susceptibility is. So the relaxation time has a strong influence on the third-order susceptibility.
\subsection{Oscillator strength}
In this section, we will discuss effecs of pseudopotential and magnetic flux quantum on the oscillator strength. The Aharonov-Bohm effect is, quite generally, a non-local effect in which a physical object travels along a closed loop through a gauge field-
free region and thereby undergoes a physical change, but we set the magnetic field value to $5T$ in this section, which has no influence on behavior that Oscillator strength versus magnetic flux $\alpha$. In Fig. 12, we presented the oscillator strength as a function of magnetic flux $\alpha$, magnetic field $\mathbf{B}$, zero point of the pseudoharmonic potential $r_{0}$ and potential strength of two-dimensional electron gas $V_{0}$. From the Fig. 12, we can see that the oscillator strength has a continuous increase until the magnetic flux comes up to 0.3 where the oscillator strength reaches the maximum value, as the magnetic flux increases. In Fig. 12a, we find that all three curves with different magnetic field overlap. So the magnetic field has no influence on the curve of oscillator strength with magnetic flux.  But in Fig. 12b and Fig. 12c the increasing $r_{0}$ and $V_{0}$ enhance the magnitude of oscillator strength. And we should note that the varying $r_{0}$ has more effect on the curve curvature of oscillator strength with magnetic flux than $V_{0}$ does. We can interpret this behavior as follows. Confinement potential is enhanced by increasing $r_{0}$ and $V_{0}$, increasing confinement potential leads to localization of wave function and reduces the transition probability between the initial state and the final state. In addition, we also observe that the magnitude of oscillator strength is very small. This indicates the transition probability is also very small. So it is very difficult to observe this transition under this confinement.
\subsection{Linear and nonlinear Optical absorption coefficient and refractive index changes}

From previous literatures [62,66], we know that the smaller magnetic field can greatly affect Linear and nonlinear optical absorption coefficient and refractive index changes of two-dimensional quantum system with pseudopotential. In this section we will report magnetic flux effect on Linear and nonlinear Optical absorption coefficient and refractive index changes. It is well-known that the magnetic flux influences the behavior of carrier wave function [47]. As we predict, in Fig. 13, the resonant peak value of linear, non-linear and total absorption coefficient show the tendency to vary periodically with magnetic flux. Also, the resonant peak is moved to higher energies due to increasing magnetic flux. This phenomenon occurs when the phase of carrier wave function is periodically changed by magnetic flux. In order to further study the effect of magnetic flux on the liner and non-linear optical properties of two-dimensional quantum ring, in Fig. 14, we plotted the curve of linear, non-linear and total refractive index changes versus photon energy with different magnetic flux value. the resonant peak value of the linear, non-linear and total refractive index changes, however, doesn't vary periodically as we expect. It is found that the increasing magnetic flux causes a continuous decrease in the magnitude of resonant peak.

\section{\textbf{Summary}}

 We have investigated the effects of an external magnetic field, magnetic flux and confinement potential on optical properties of a two-dimensional quantum ring. Our results shown: (i) The resonance peak of the third-order susceptibility decreases as the magnetic field increases, and that the peak position moves towards higher energies. Also, the external magnetic field has a weak influence on the third-order susceptibility in two-dimensional quantum ring. (ii) The resonance peak of third-order susceptibility increases dramatically as $r_{0}$ increases. It is also found that the resonance peak has a red shift.
(iii) Increasing the potential $V_{0}$ leads to the increment in resonance peak of third-order susceptibility, meanwhile, we can also see that the peak position shifts towards higher energies as $V_{0}$ increases. (iv) Aluminium concentration plays an important role in the third-order susceptibility of two-dimensional $Al_{x}Ga_{1-x}As$ pseudodot system. (v) The relaxation time has a strong influence on the third-order susceptibility. (vi) Unlike magnetic field, the magnetic flux, potential $V_{0}$, zero point $r_{0}$, have a great influence on the oscillator strength. (vii) Resonant peak value of the linear, non-linear and total absorption coefficient varies periodically with magnetic flux, while not for refractive index changes.

  In conclusion, optical properties of two-dimensional quantum ring are strongly affected by the external magnetic field, confinement potential, magnetic flux, aluminium concentration and relaxation time. Especially for the effect of magnetic flux on the optical properties, the researcher should take into account in designing optical devices. Finally we hope our research can contribute to understanding the two-dimensional quantum ring with pseudopotential better.
\section{\textbf{Acknowledgement}}
We are indebted to Dr. Ghasem Rezaei for fruitful discussions. This work is supported by National Natural Science Foundation of China(under
Grant No. 11074055).

\newpage

\section{caption}

Fig. 1. The third-order susceptibility as a function of the photon energy with $r_{0} = 4 nm$ and $V_{0} = 350 meV$ for three different magnetic field values.

Fig. 2. The energy differences as a function of the external magnetic field with $r_{0} = 4 nm$ and $V_{0} = 350 meV$.

Fig. 3. The product of geometric factor as a function of the external magnetic field with $r_{0} = 4 nm$ and $V_{0} = 350 meV$.

Fig. 4. The third-order susceptibility as a function of the photon energy with $B = 1 T$ and $V_{0} = 350 meV$ for three different $r_{0}$ values.

Fig. 5. The energy differences as a function of the geometric size of dot $r_{0}$ with $B = 1 T$ and $V_{0} = 350 meV$.

Fig. 6. The product of geometric factor as a function of the geometric size of dot $r_{0}$ with $B = 1 T$ and $V_{0} = 350 meV$.

Fig. 7. The third-order susceptibility as a function of the photon energy with $B = 1 T$ and $r_{0} = 4 nm$ for three different $V_{0}$ values.

Fig. 8. The energy differences as a function of chemical potential $V_{0}$ with $B = 1 T$ and $r_{0} = 4 nm$.

Fig. 9. The product of geometric factor as a function of chemical potential $V_{0}$ with $B = 1 T$ and $r_{0} = 4 nm$.

Fig. 10. The third-order susceptibility of two-dimensional $Al_{x}Ga_{1-x}As$ quantum system as a function of the photon energy with $B = 1 T$, $r_{0} = 4 nm$ and $V_{0} = 350 meV$ for three different aluminium concentration $x$ values.

Fig. 11. The third-order susceptibility as a function of the photon energy with $B = 1 T$, $r_{0} = 4 nm$ and $V_{0} = 350 meV$ for three different $\tau$ (relaxation time) values.

Fig. 12. (a) The oscillator strength versus magnetic flux with $r_{0} = 2 nm$ and $V_{0} = 100 meV$ for three different magnetic field values $B$. (b) The oscillator strength versus magnetic flux with $r_{0} = 2 nm$ and $B=5T$ for three different chemical potential values$V_{0}$. (c) The oscillator strength versus magnetic flux with $V_{0} = 100 meV$ and $B=5T$ for three different zero point values $r_{0}$.

Fig. 13. The linear, non-linear and total absorption coefficient as a function of magnetic flux with $r_{0} = 2 nm$ and $V_{0} = 100 meV$ and $I=0.4MW/cm^{2}$.
Fig. 14. The linear, non-linear and total refractive index changes as a function of magnetic flux with $r_{0} = 2 nm$, $V_{0} = 100 meV$ and $I=0.4MW/cm^{2}$.

\newpage
\begin{figure}[tbp]
\begin{center}
\includegraphics[scale=1.2]{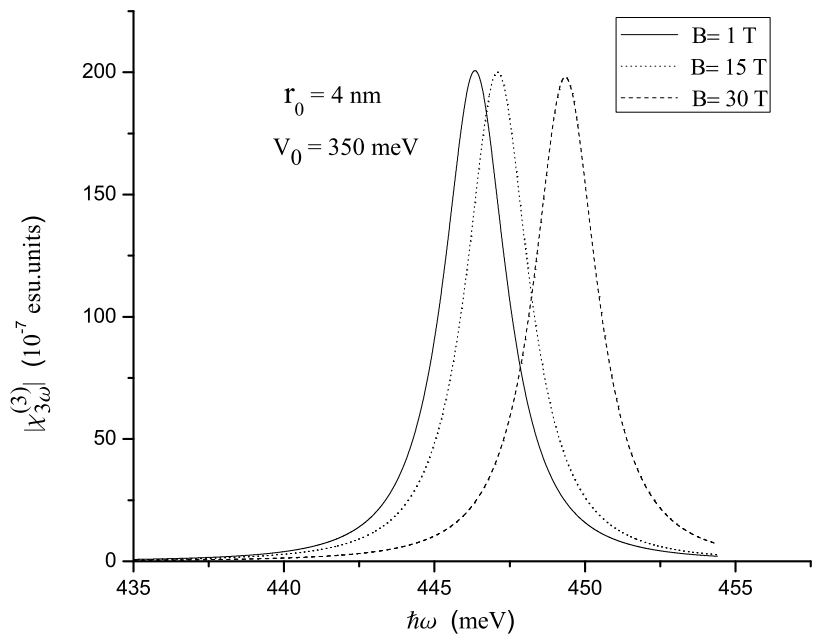}
\end{center}
\caption{The third-order susceptibility as a function of the photon energy with $r_{0} = 4 nm$ and $V_{0} = 350 meV$ for three different magnetic field values.}
\end{figure}

\begin{figure}[tbp]
\begin{center}
\includegraphics[scale=1.2]{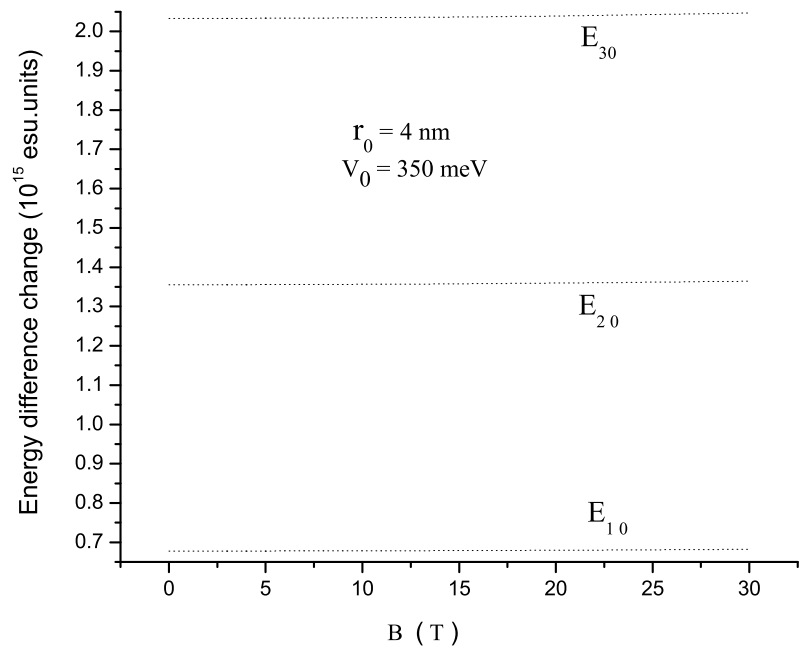}
\end{center}
\caption{The energy differences as a function of the external magnetic field with $r_{0} = 4 nm$ and $V_{0} = 350 meV$.}
\end{figure}

\begin{figure}[tbp]
\begin{center}
\includegraphics[scale=1.2]{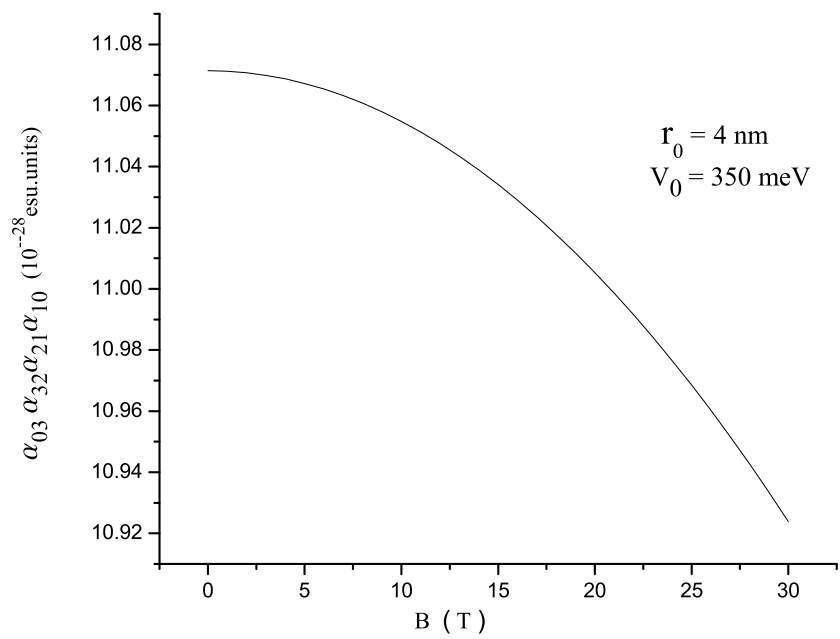}
\end{center}
\caption{The product of geometric factor as a function of the external magnetic field with $r_{0} = 4 nm$ and $V_{0} = 350 meV$.}
\end{figure}

\begin{figure}[tbp]
\begin{center}
\includegraphics[scale=1.2]{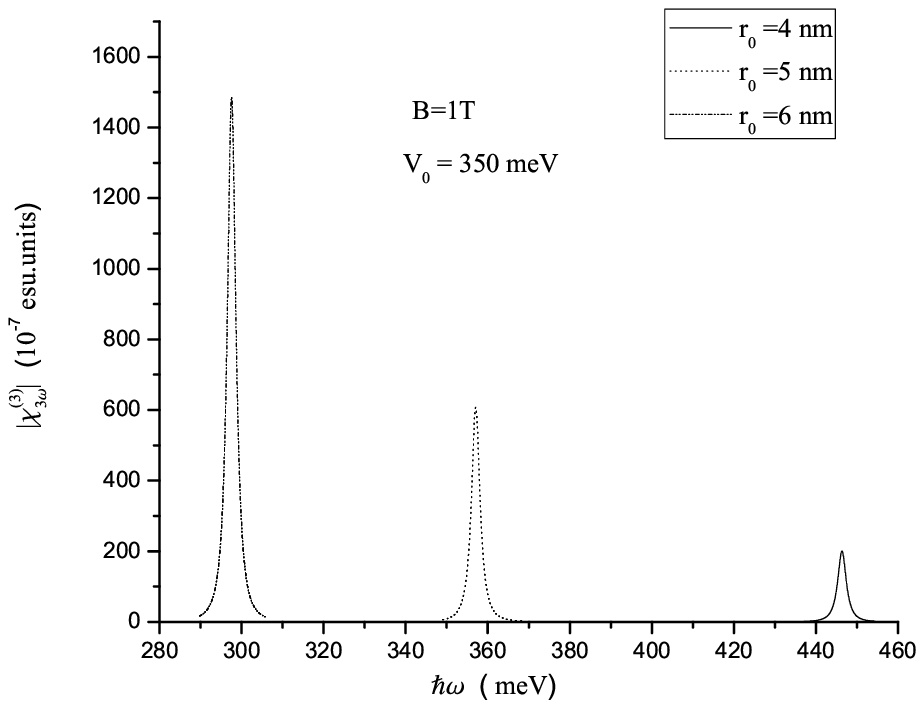}
\end{center}
\caption{Third-order susceptibility as a function of the photon energy with $B = 1 T$ and $V_{0} = 350 meV$ for three different $r_{0}$ values.}
\end{figure}
\begin{figure}[tbp]
\begin{center}
\includegraphics[scale=1.2]{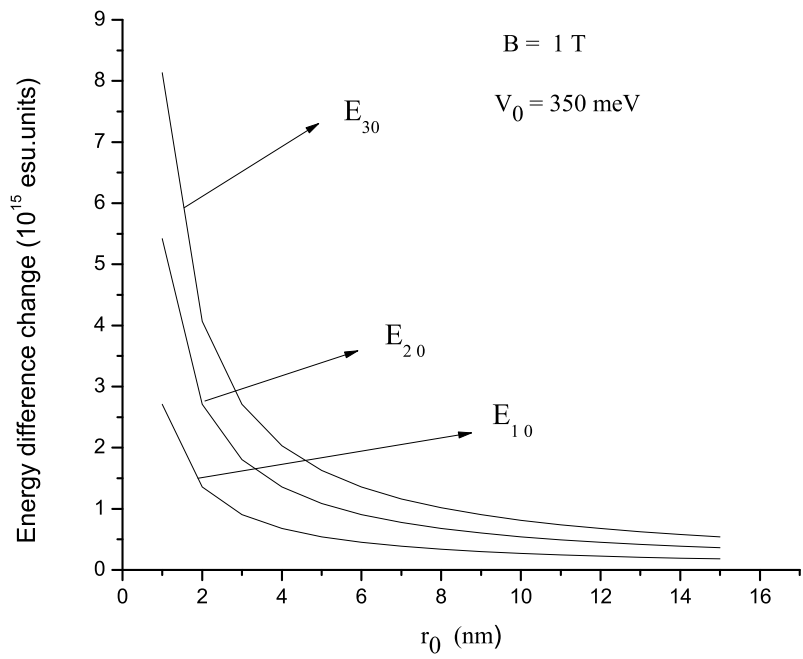}
\end{center}
\caption{The energy differences as a function of the geometric size of dot $r_{0}$ with $B = 1 T$ and $V_{0} = 350 meV$.}
\end{figure}
\begin{figure}[tbp]
\begin{center}
\includegraphics[scale=1.2]{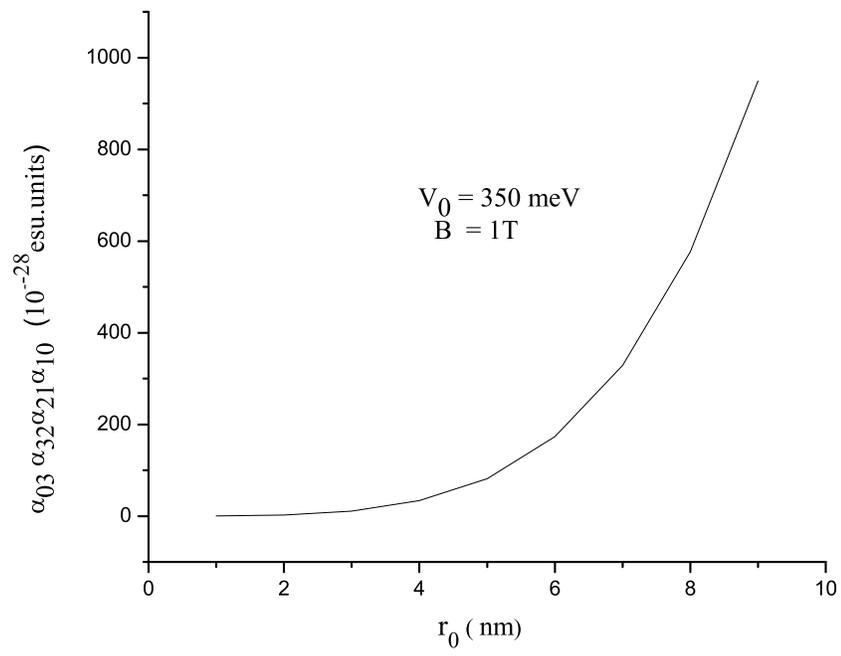}
\end{center}
\caption{The product of geometric factor as a function of the geometric size of dot $r_{0}$ with $B = 1 T$ and $V_{0} = 350 meV$.}
\end{figure}
\begin{figure}[tbp]
\begin{center}
\includegraphics[scale=1.2]{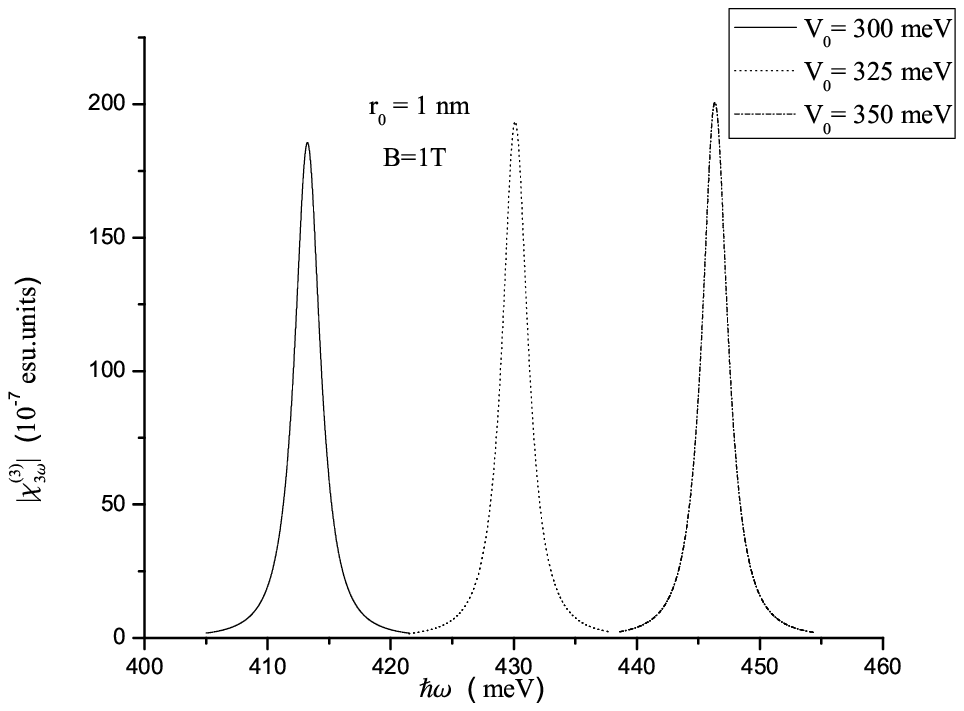}
\end{center}
\caption{ The third-order susceptibility as a function of the photon energy with $B = 1 T$ and $r_{0} = 4 nm$ for three different $V_{0}$ values.}
\end{figure}
\begin{figure}[tbp]
\begin{center}
\includegraphics[scale=1.2]{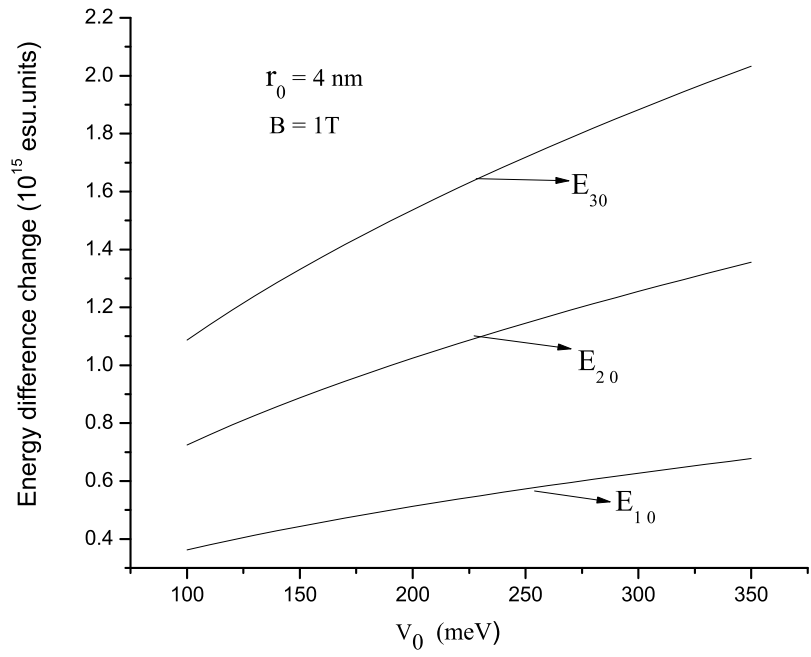}
\end{center}
\caption{The energy differences as a function of chemical potential $V_{0}$ with $B = 1 T$ and $r_{0} = 4 nm$.}
\end{figure}
\begin{figure}[tbp]
\begin{center}
\includegraphics[scale=1.2]{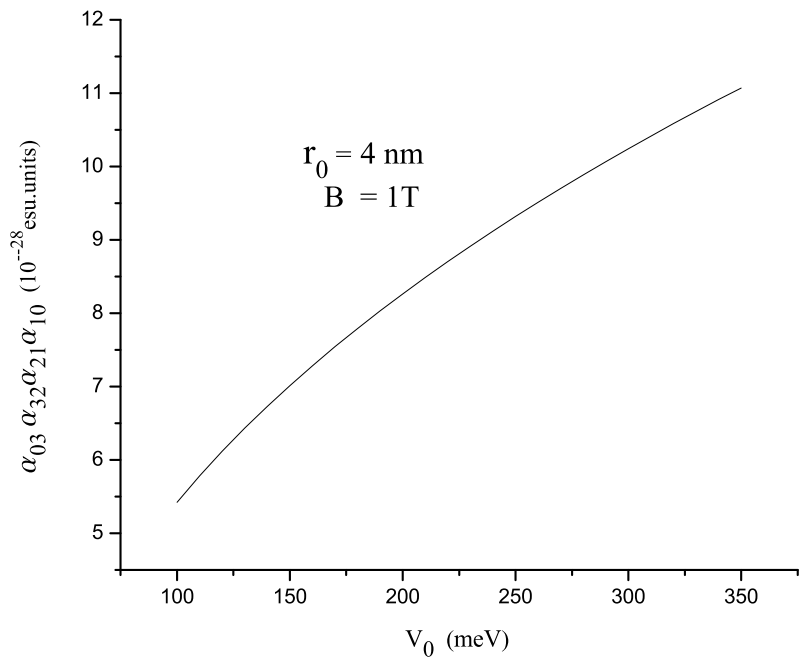}
\end{center}
\caption{The product of geometric factor as a function of chemical potential $V_{0}$ with $B = 1 T$ and $r_{0} = 4 nm$.}
\end{figure}
\begin{figure}[tbp]
\begin{center}
\includegraphics[scale=1.2]{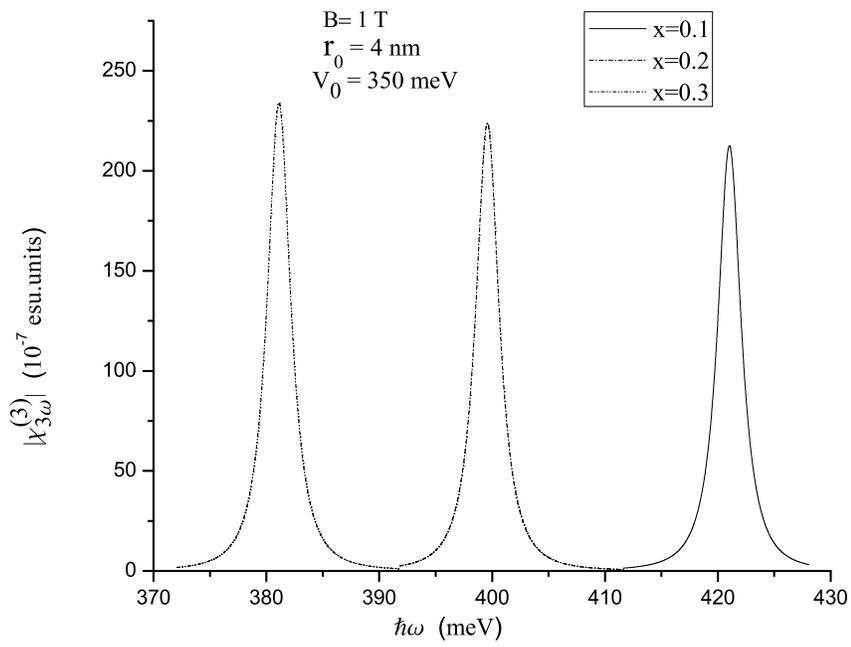}
\end{center}
\caption{The third-order susceptibility of two-dimensional $Al_{x}Ga_{1-x}As$ pseudodot system as a function of the photon energy with $B = 1 T$, $r_{0} = 4 nm$ and $V_{0} = 350 meV$ for three different aluminium concentration $x$ values.}
\end{figure}
\begin{figure}[tbp]
\begin{center}
\includegraphics[scale=1.2]{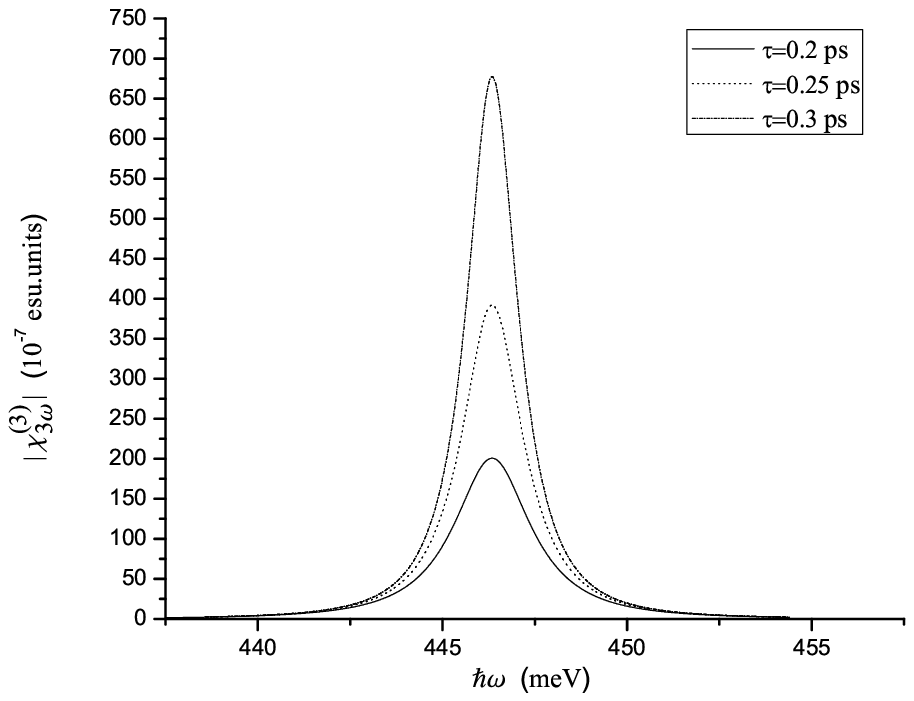}
\end{center}
\caption{The third-order susceptibility as a function of the photon energy with $B = 1 T$, $r_{0} = 4 nm$ and $V_{0} = 350 meV$ for three different $\tau$ (relaxation time) values.}
\end{figure}
\begin{figure}[tbp]
\begin{center}
\includegraphics[scale=1.2]{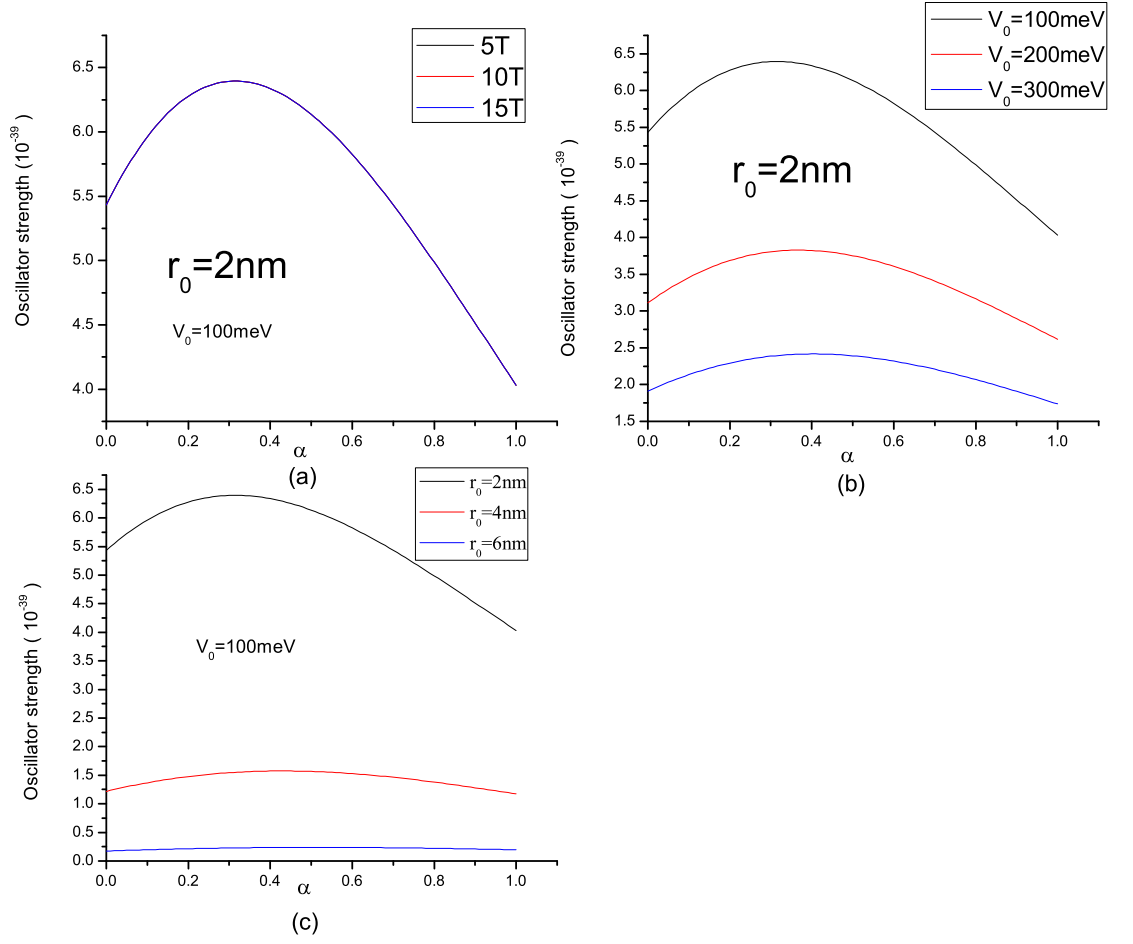}
\end{center}
\caption{(a) The oscillator strength versus magnetic flux with $r_{0} = 2 nm$ and $V_{0} = 100 meV$ for three different magnetic field values $B$. (b) The oscillator strength versus magnetic flux with $r_{0} = 2 nm$ and $B=5T$ for three different chemical potential values$V_{0}$. (c) The oscillator strength versus magnetic flux with $V_{0} = 100 meV$ and $B=5T$ for three different zero point values $r_{0}$.}
\end{figure}
\begin{figure}[tbp]
\begin{center}
\includegraphics[scale=1.2]{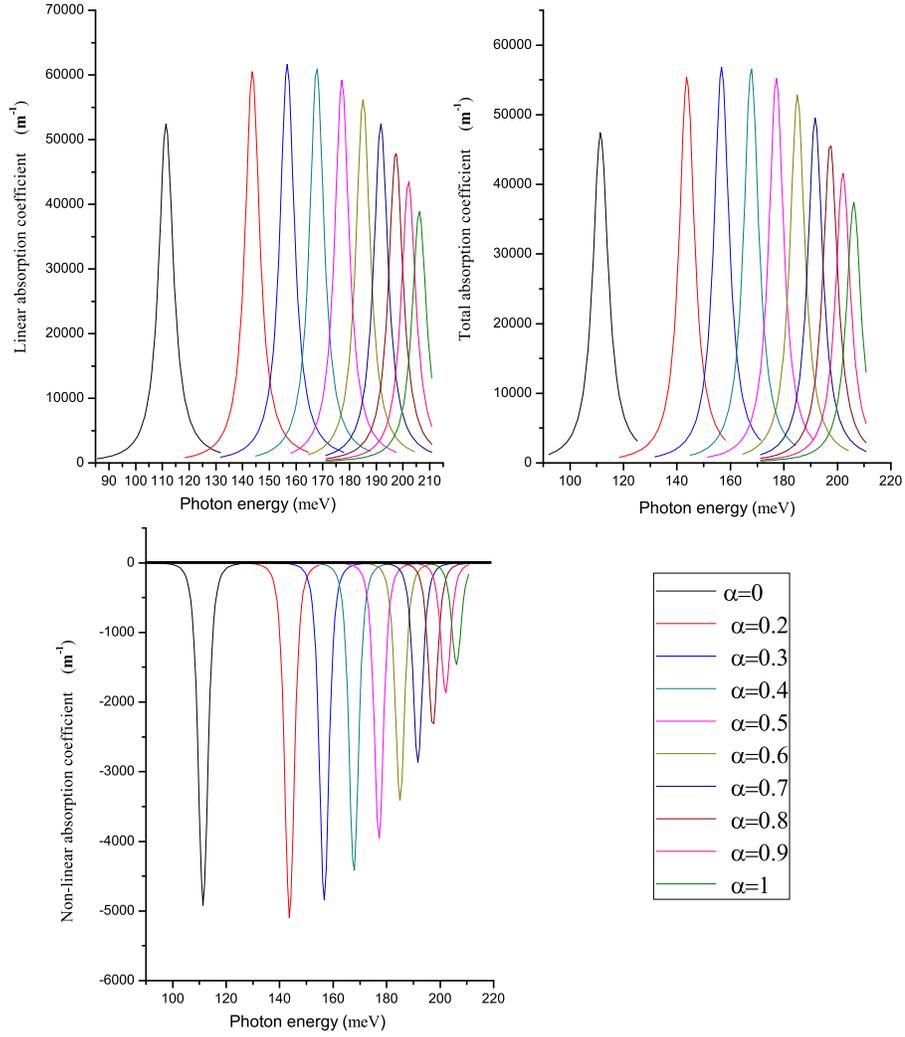}
\end{center}
\caption{The linear, non-linear and total absorption coefficient as a function of magnetic flux with $r_{0} = 2 nm$ and $V_{0} = 100 meV$ and $I=0.4MW/cm^{2}$.}
\end{figure}
\begin{figure}[tbp]
\begin{center}
\includegraphics[scale=1.2]{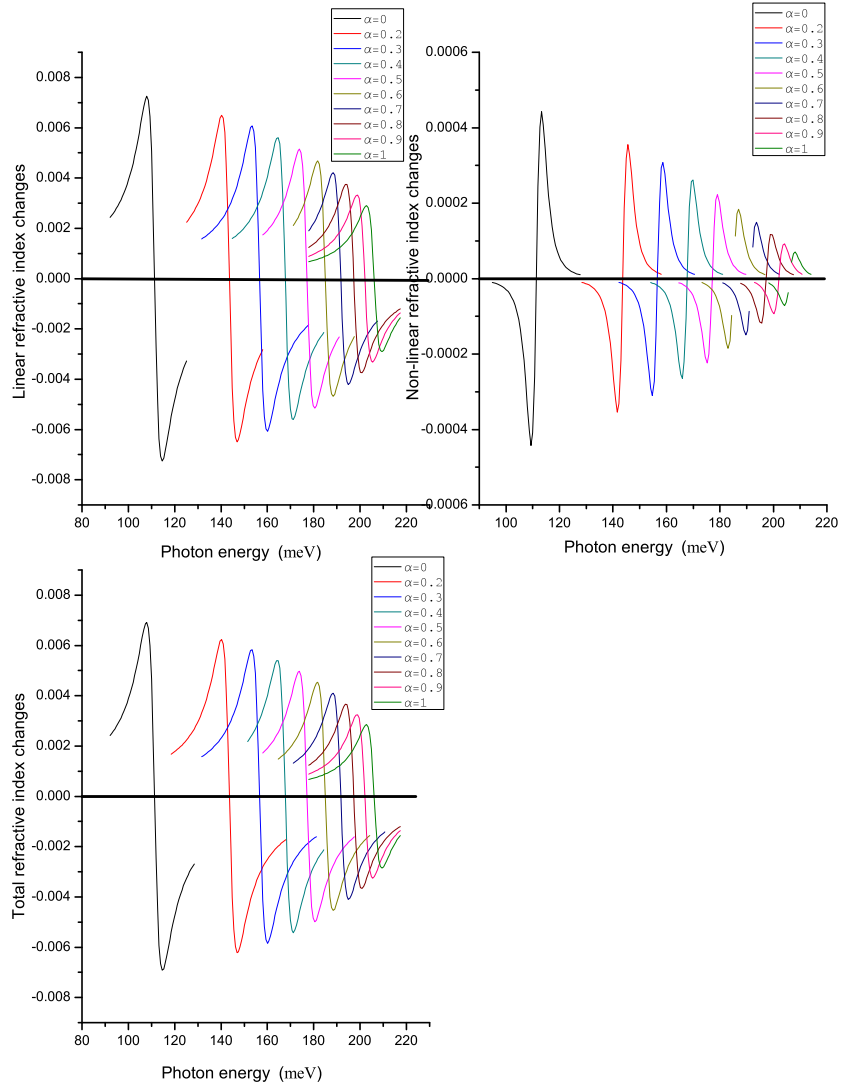}
\end{center}
\caption{ The linear, non-linear and total refractive index changes as a function of magnetic flux with $r_{0} = 2 nm$, $V_{0} = 100 meV$ and $I=0.4MW/cm^{2}$.}
\end{figure}
\end{document}